\documentclass{article}
\usepackage[dvips]{graphicx}
\usepackage{amssymb}
\usepackage{amsmath}
\usepackage{amscd}
\usepackage{mathrsfs}
\usepackage{latexsym}
\usepackage{makeidx}

\newtheorem{theorem}{Theorem}
\newtheorem{definition}{Definition}

\newtheorem{proposition}{Proposition}

\newcommand{\bd}{\begin{definition}}
\newcommand{\ed}{\end{definition}}
\newcommand{\bp}{\begin{proposition}}
\newcommand{\be}{\begin{equation}}
\newcommand{\ee}{\end{equation}}
\newcommand{\bea}{\begin{eqnarray}}
\newcommand{\eea}{\end{eqnarray}}
\newcommand{\ba}{\begin{array}}
\newcommand{\ea}{\end{array}}

\date{}
\author{Claudio Garola\footnote{Dipartimento di Fisica and Sezione INFN; Universit\`a di Lecce, via Arnesano, 73100 Lecce. E-mail address: garola@le.infn.it.} \, and \, Sandro Sozzo\footnote{Dipartimento di Fisica and Sezione INFN; Universit\`a di Lecce, via Arnesano, 73100 Lecce. E-mail address: sozzo@le.infn.it.}}

\title{\Large{\textbf{On the Physical Interpretation of Partial Traces: Two Nonstandard Viewpoints}}}
\begin{document}
\maketitle

\begin{abstract}
\noindent
Mixed states are introduced in physics in order to express our ignorance about the actual state of a physical system and are represented in standard quantum mechanics (QM) by density operators. Such operators also appear if one considers a (pure) entangled state of a compound system $\Omega$ and performs partial traces on the projection operator representing it. Yet, they do not represent mixed states (or \emph{proper mixtures}) of the subsystems in this case, but \emph{improper mixtures}, since the coefficients in the convex sums expressing them never bear the ignorance interpretation. Hence, one cannot attribute states to the subsystems of a compound physical system in QM (\emph{subentity problem}). We discuss here two alternative proposals that can be worked out within the \emph{Brussels} and \emph{Lecce} approaches. We firstly summarize the general framework provided by the former, which suggests that improper mixtures could be considered as new pure states. Then, we show that improper mixtures can be considered as true (yet nonpure) states also according to the latter. The two proposals seem to be compatible notwithstanding their different terminologies.

\vspace{.3cm}
\end{abstract}

\section{Introduction}
In the standard formulation of quantum mechanics (QM) a physical system $\Omega$ is associated with a separable complex Hilbert space $\mathcal H$ and the states of $\Omega$ are represented by density operators on $\mathcal H$, which reduce to one--dimensional (orthogonal) projection operators in the case of \emph{pure} states. Every density operator $W$ representing a \emph{mixed} state, or \emph{proper mixture}, $S$ of $\Omega$ can be expressed in many ways as a convex combination of pure states, and a decomposition $W=\sum_{i} p_{i}| \psi_{i} \rangle\langle \psi_{i}|$ exists in which every coefficient $p_i$ denotes the probability that $\Omega$ is in the state $S_i$ represented by the projection operator $| \psi_{i} \rangle\langle \psi_{i}|$. This probability expresses our ignorance about the real state of $\Omega$, hence also about the result of a measurement testing whether the property $E_i$ of $\Omega$ represented by $|\psi_{i}\rangle\langle\psi_{i}|$ is possessed by $\Omega$. Yet, if $\langle\psi_i|\psi_j \rangle=\delta_{ij}$, every $E_i$ is \emph{objective} in $S$, in the sense that it can be considered as either possessed or not possessed by $\Omega$ independently of any measurement. Indeed, the former case occurs if the real state of $\Omega$ is $S_i$, the latter case if the real state is $S_j$, with $j \ne i$. Now, let $\Omega$ be a compound system, made up by two subsystems $\Omega_1$ and $\Omega_2$, in a pure entangled state $S_P$ represented by the projection operator $| \psi \rangle\langle \psi|$. Let $|\psi\rangle=\sum_{i}\sqrt{p_i}|\phi_{i}(1)\rangle |\chi_{i}(2)\rangle$ (where $0 \ne p_i$, and $p_i \ne 1$ since $S_P$ is entangled) be the \emph{biorthogonal decomposition} of $|\psi\rangle$, so that $|\psi\rangle\langle\psi|=\sum_{i,j}\sqrt{p_{i}p_{j}}|\phi_{i}(1)\rangle\langle\phi_{j}(1)| \otimes |\chi_{i}(2)\rangle\langle\chi_{j}(2)|$. If one considers $\Omega_1$ only, the physical information provided by QM on it can be attained by performing the partial trace of $|\psi\rangle\langle\psi|$ with respect to $\Omega_2$, thus getting $W_1= Tr_{2} |\psi \rangle\langle \psi|=\sum_{i} p_{i} |\phi_{i}(1)\rangle\langle\phi_{i}(1)|$, and applying standard quantum rules. The density operator $W_1$ is formally similar to $W$. Yet, a coefficient $p_i$ in it denotes the probability of actualizing the property $E_i(1)$ of $\Omega$ represented by the projection operator $|\phi_{i}(1)\rangle\langle\phi_{i}(1)| \otimes I_2$ whenever a measurement occurs (or the transition probability from $S_P$ to the pure state represented by $|\phi_{i}(1)\rangle\langle\phi_i(1)| \otimes |\chi_{i}(2)\rangle\langle\chi_i(2)|$ if the measurement is ideal), but it cannot denote the probability that $\Omega_1$ actually is in the state $S_{i}(1)$ represented by $|\phi_{i}(1)\rangle\langle\phi_{i}(1)|$. Indeed, $E_i(1)$ should then be objective, as the property $E_i$ considered above, while it is \emph{nonobjective} in $S_P$ according to the standard interpretation of QM (that is, one cannot consider $E(1)$ as either possessed or not possessed by $\Omega$ in the state $S_P$ if a measurement is not performed).\footnote{Nonobjectivity is commonly believed to be an intrinsic and uneliminable feature of QM because of some mathematical results, as the \emph{Bell--Kochen--Specker theorem} (that proves the \emph{contextuality} of QM, which means that the result of the measurement of a property of a physical system in a given state is not prefixed, but depends on the measurement context) \cite{b66,ks67}, and the \emph{Bell theorem} (that proves the \emph{nonlocality} of QM, which means that contextuality occurs also at a distance) \cite{b64}. Yet, it is the deep root of most problems that afflict the standard interpretation of QM and generates a lot of paradoxes and conceptual difficulties (in particular, the \emph{objectification problem} in the quantum theory of measurement, see, \emph{e.g.}, \cite{blm91,bs96}). \label{nogo}} Basing on this conclusion, one can show that no decomposition of $W_1$ bears the above \emph{ignorance interpretation}. Hence, some authors say that $W_1$ represents an \emph{improper mixture}, distinguishing it from a proper mixture as $W$ (see, \emph{e.g.}, \cite{hk69}--\cite{blm91}).

It follows from the above reasonings that the density operators obtained by performing partial traces generally neither represent pure nor mixed states of the component subsystems in standard QM, so that these subsystems can never be considered as independent entities, which raises the so--called \emph{subentity problem}. In particular, two distinct approaches to the foundations of QM are directly or indirectly concerned with the solution of this problem, namely, the Brussels and the Lecce approach. Indeed, the former (briefly, \emph{BR approach}), elaborated by Aerts and his collaborators in Brussels in the last two decades, continues the work started by Jauch and Piron in the sixties and the seventies in Geneva \cite{j68,p76}, with the aim of providing not only a physical justification of the mathematical apparatus of QM by means of an operational foundation of this theory, but also a better description of compound entities by introducing some changes in the theory itself. The Lecce approach, instead, basing on a criticism of the theorems that are maintained to prove the nonobjectivity of the theory, offers a non--standard interpretation of QM (\emph{semantic realism}, or \emph{SR}, interpretation) that is local and noncontextual (it preserves anyway the mathematical apparatus of QM and its statistical interpretation) with the aim of avoiding the paradoxes that afflict standard QM. Then, this framework provides also the basic elements for a solution of the subentity problem.

We note that the foregoing approaches have a number of common features, and seem to come to the same conclusions about many problematical issues of QM \cite{gps06,santesi}. In particular, both the BR and the Lecce approach suggest that a broader theory can exist that embodies QM but says more than it. Here, however, we are only concerned with a comparison of the nonstandard proposals emerging from the two approaches for a solution of the subentity problem. To this end, we resume in Sec. 2 the essentials of the BR approach, and discuss in Sec. 3 a possible solution of the subentity problem within this approach, which may entail a breakdown of the linearity of QM. Then we briefly present the Lecce approach in Sec. 4, and show in Sec. 5 that a solution of the subentity problem can be worked out also within this approach. Both solutions propose considering improper mixtures as new states of the physical system, hence they seem to be compatible, despite their different terminologies.

\section{The Brussels approach}
We supply in this section a summary of the basic features of the BR approach that are required for discussing a solution of the subentity problem \cite{a81}--\cite{a02b} propounded by this approach. Our presentation is synthetic and partially original, hence possible defects or misunderstandings must be charged to ourselves.

Within the BR approach, the term \emph{physical entity} is preferred to the standard term \emph{physical system} used so far. Then, let $\Omega$ be a physical entity. According to the BR approach, $\Omega$ is a \emph{state property entity} iff it is associated with a set of states $\Sigma$ and a set of properties $\cal L$. A state $p$ of $\Omega$ intuitively represents a ``mode of being'' of $\Omega$. This means that at each moment the entity $\Omega$ ``is'' in a specific state $p \in \Sigma$. A property $a$ of $\Omega$ is an attribute of $\Omega$. The property $a$ can be \emph{actual} in the state $p$, which means that $p$ is such that $\Omega$ has (or possesses) the property $a$ ``in acto'', or \emph{potential}, which means that $p$ is such that $\Omega$ does not possess the property $a$, but can acquire it. Hence,  a mapping $\rho: \Sigma \times {\cal L} \rightarrow \{ A, P \}$ (where $A$ stands for \emph{actual} and $P$ for \emph{potential}) can be introduced. Thus, if the entity $\Omega$ is in a state $p$, one can consider the set $\xi(p)= \{ a \in {\cal L} \ | \ \rho(p,a)=A \}$ of all properties that are actual in $p$. Analogously, for a property $a$ of $\Omega$, one can consider the set $\kappa(a)= \{ p \in \Sigma \ | \   \rho(p,a)=A \}$ of all states that make $a$ actual. Hence, for every $p \in \Sigma$ and $a \in \cal L$, $\rho(p,a)=A$ iff $a \in \xi(p)$, or $p \in k(a)$. 

Let $\Omega$ be a state property entity and let $a,b \in {\cal L}$. If, for every $p \in \Sigma$, $ \rho(p,a)=A$ implies $\rho(p,b)=A$, one says that $a$ \emph{implies} $b$ and writes $a < b$ (hence $a < b$ iff $\kappa(a) \subseteq \kappa(b)$). It is easy to see that $<$ is a preorder relation on ${\cal L}$. Analogously, let $p, q \in \Sigma$. If for every $a \in {\cal L}$, $\rho(q,a)=A$ implies $\rho(p,a)=A$, then one says that $p$ \emph{ property implies} $q$ and writes $p < q$ (hence $p < q$ iff $\xi(q) \subseteq \xi(p)$). It is easy to see that $<$ is a preorder relation on $\Sigma$.   

The above remarks suggest introducing the following definition (where, for every set $\cal K$, ${\cal P}({\cal K})$ denotes the power set of $\cal K$).

\begin{definition}
Let $\Omega$ be a state property entity characterized by a set of \emph{states} $\Sigma$, a set of \emph{properties} ${\cal L}$, and two mappings $\xi : p \in \Sigma \rightarrow  \xi(p) \in {\cal P}({\cal L})$ and $\kappa : a \in {\cal L} \rightarrow \kappa(a) \in {\cal P}(\Sigma)$ (where $\xi(p)$ denotes the set of properties that are \emph{actual} if the entity $\Omega$ is in the state $p$ and $\kappa(a)$ the set of states in which the property $a$ is actual). For every $p,q \in \Sigma$ and $a,b \in {\cal L}$, we put $p < q$ iff  $\xi(q) \subseteq \xi(p)$, $a < b$ iff $\kappa(a) \subseteq \kappa(b)$, and say that $\Omega$ is described by the structure $(\Sigma, <, {\cal L}, <, \xi, \kappa)$ (briefly, $(\Sigma, {\cal L}, \xi, \kappa)$). Furthermore, we say that $(\Sigma, {\cal L}, \xi, \kappa)$ is a \emph{state property system} iff $({\cal L}, <)$ is a complete lattice (whose meet and join we denote by $\wedge$ and $\vee$, respectively), and the following conditions hold for every $p \in \Sigma$.

(i) Let $I$ and $0$ be the maximal and minimal elements of $({\cal L}, <)$, respectively. Then, $I \in \xi(p)$ and $0 \not\in \xi(p)$. 

(ii) Let $\{ a_i \in {\cal L} \}_{i \in {\cal I}}$ be a family of elements of $\cal L$. Then, $\wedge_ia_i \in \xi(p)$ iff, for every $i \in {\cal I}$, $a_i \in \xi(p)$.
\end{definition}

Following Aerts, we suppose from now on that every physical entity $\Omega$ considered here is described by a state property system. One can then prove that the standard formalism of QM can be recovered by adding suitable axioms on the state property system $(\Sigma,{\mathcal L}, \xi, \kappa)$. We firstly collect all the required axioms in the following definition.

\begin{definition}
Let $(\Sigma, {\cal L}, \xi, \kappa)$ be a state property system, and let $\cal A$ be the set of all atoms of the lattice $({\cal L},<)$.

(i) $(\Sigma, {\cal L}, \xi, \kappa)$ satisfies the \emph{state determination axiom} if, for every $p, q \in \Sigma$,
$\bigwedge_{a \in  \xi(p)}a = \bigwedge_{b \in \xi(q)}b$ implies $p=q$.

(ii) $(\Sigma, {\cal L}, \xi, \kappa)$ satisfies the \emph{atomicity axiom} if, for every $p \in \Sigma$, the element $\bigwedge_{a \in \xi(p)}a$ is an atom of $({\cal L},<)$. 

(iii) $(\Sigma, {\cal L}, \xi, \kappa)$ satisfies the \emph{orthocomplementation axiom} if $({\cal L},<)$ is \emph{orthocomplemented}, that is, there exists a mapping $' :{\cal L} \rightarrow {\cal L}$ such that, for every $a, b \in {\cal L}$, $(a')'= a$, $a < b$ implies $b' < a'$, $a \wedge a' = 0$ and $a \vee a' = I$.

(iv) $(\Sigma, {\cal L}, \xi, \kappa)$ satisfies the \emph{covering law axiom} if, for every $a,x \in \cal L$, and $b \in \cal A$, $a < x < a \vee b$ implies either $x=a$ or $x= a \vee b$.

(v) $(\Sigma, {\cal L}, \xi, \kappa)$ satisfies the \emph{weak modularity axiom} if $({\cal L}, <)$ is orthocomplemented and, for every $a,b \in \cal L$, $a< b$ implies $(b \wedge a') \vee a =b$.

(vi) $(\Sigma, {\cal L}, \xi, \kappa)$ satisfies the \emph{plane transitivity axiom} if, for every $s, t \in \cal A$, there exist two distinct atoms $s_1, s_2 \in \cal A$ and an automorphism $f$ on $\cal L$ such that $t=f(s)$ and, for every $a \in [0, s_1 \vee s_2]$, $f(a)=a$. 

(vii) $(\Sigma, {\cal L}, \xi, \kappa)$ satisfies the \emph{irreducibility axiom} if $({\cal L}, <)$ is orthocomplemented and, for every $b \in \cal L$, $b = (b \wedge a) \vee (b \wedge a')$ $\forall a \in \cal L$, implies either $b=0$ or $b=I$.

(viii) $(\Sigma, {\cal L}, \xi, \kappa)$ satisfies the \emph{infinite length axiom} if $({\cal L}, <)$ is orthocomplemented and contains an infinite set of \emph{mutually orthogonal} elements (where $b,c \in \cal L$ are mutually orthogonal iff a property $a \in \cal L$ exists such that $b < a$ and $c < a'$).
\end{definition}

The following representation theorem can now be proved.

\begin{theorem} 
Let $\Omega$ be a physical entity described by the state property system $(\Sigma, {\cal L}, \xi, \kappa)$ satisfying all the axioms in Definition 2. Then, $({\cal L}, <)$ is isomorphic to the complete lattice $({\cal L}({\cal H}), \subseteq)$ of the projection operators on an infinite dimensional real, complex or quaternionic Hilbert space ${\cal H}$. The atoms of $({\cal L}({\cal H}), \subseteq)$ (\emph{i.e.}, the one--dimensional projection operators on $\cal H$) are in a one--to--one correspondence with the atoms of $({\cal L},<)$ and with the elements of $\Sigma$.\footnote{Note that the existence of a one--to--one mapping of $\Sigma$ onto $\cal A$ follows from the atomicity axiom. This axiom is sometimes replaced by a weaker axiom within the BR approach (see, \emph{e.g.}, \cite{a99b}), so that the set $\Sigma$ must be substituted by the subset $\Lambda$ of all atoms of $(\Sigma, <)$ (or \emph{atomic states}) in Theorem 1. We have chosen here the stronger statement for the sake of simplicity.} The orthocomplementation is induced by the orthogonality structure of ${\cal H}$.
\end{theorem}

The representation theorem stated above makes it possible to recover the standard quantum formalism from the operational structure of state property system introduced by the BR approach. The problem of proving it engaged a number of scholars, and its solution gave an end to the so--called \emph{coordinatization problem} \cite{bc81}. The proof is long and complicated and requires some important mathematical results, such as the Piron representation theorem \cite{p76}, the fundamental theorem of projective geometry \cite{v68}, the Sol\`{e}r theorem \cite{s95} and the Gleason theorem \cite{g57}. The final result is relevant, since it provides a (partial) operational justification of the standard formalism of QM, which thus seems more firmly founded. But, then, a new problem occurs. Indeed, it is well known that QM meets some difficulties whenever one wants to apply it to compound entities. Therefore, we dedicate the next section to this issue.

\section{The subentity problem within the Brussels approach}
In order to discuss the subentity problem we must preliminary introduce a remark on the representation theorem stated in Sec. 2. To be precise, we stress that this theorem establishes a one--to--one correspondence between the atoms of $({\cal L}({\cal H}), \subseteq)$ and the elements of $\Sigma$, which implies that $\Sigma$ does not contain mixed states. Hence, all axioms in Definition 2 refer to a state property system for which only pure states are considered. This is consistent with the intuitive notion of state as a ``mode of being'' of the entity $\Omega$. According to the BR approach, the reality of a quantum entity in QM is expressed by a pure state, represented by a one--dimensional projection operator of the corresponding Hilbert space. Proper mixtures have obviously been taken into account within the BR approach, but they are not classified as modes of being of the entity, since they do not represent the reality of the entity but a lack of knowledge about this reality. Indeed, the probabilities that appear in them express (as in the standard interpretation of QM, see Sec. 1) our subjective ignorance about the actual state of the entity \cite{a99a}.

Let us come to the definition of subentity according to the BR approach.

Firstly, let $\Omega$ and $\Omega'$ be two physical entities described by the state property systems $(\Sigma, {\cal L}, \xi, \kappa)$ and $(\Sigma', {\cal L}', \xi', \kappa')$, respectively. If $\Omega$ is to be a subentity of $\Omega'$, it is reasonable to demand that, if $\Omega'$ is in a state $p'$, then $\Omega$, as part of $\Omega'$, is in a state $m(p')$. The mapping $m$ that is thus introduced should be surjective (each state of $\Omega$ corresponds to at least one state of $\Omega'$) but not necessarily injective (different states of $\Omega'$ can correspond to the same state of $\Omega$). 

Secondly, if $\Omega$ is to be a subentity of $\Omega'$, each property $a$ of $\Omega$ should correspond to a property $n(a)$ of $\Omega'$. The mapping $n$ that is thus introduced should be injective (if two properties of $\Omega$ are different, they are also different when are considered as properties of $\Omega'$) but not necessarily surjective (there are properties of $\Omega'$ that do not correspond to properties of $\Omega$). 

Finally, if the property $a$ is actual when $\Omega$ is in the state $m(p')$, then the property $n(a)$ should be actual when $\Omega'$ is in the state $p'$ (\emph{covariance principle}, see, \emph{e.g.}, \cite{a99b}). 

Because of the above arguments the BR approach provides a definition of subentity that can be summarized as follows.

\begin{definition} \label{confun} Let $\Omega$ and $\Omega'$ be two physical entities described by the state property systems $(\Sigma, {\cal L}, \xi, \kappa)$ and $(\Sigma', {\cal L}', \xi', \kappa')$, respectively. We say that $\Omega$ is a \emph{subentity} of $\Omega'$ iff there exists a surjective mapping $m : p' \in \Sigma' \rightarrow m(p') \in \Sigma$ and an injective mapping $n : a \in {\cal L} \rightarrow n(a) \in {\cal L}'$, such that, for every $p' \in \Sigma'$ and $a \in {\cal L}$, $a \in \xi(m(p'))$ iff $n(a) \in \xi'(p')$.
\end{definition}

Let us now consider QM. Because of the representation theorem, all elements of a state property system satisfying the axioms in Definition 2 can be represented within the standard mathematical apparatus of QM. In particular, the notion of subentity can be restated in QM by substituting $\Sigma$ with the set of all atoms of $({\cal L}({\cal H}), \subseteq)$, and $\cal L$ with the set of all projection operators of $({\cal L}({\cal H}), \subseteq)$. But, then, one can prove that the subsystems of a compound system (or \emph{entity} according to the terminology of the BR approach) are not subentities in the sense of Definition 3. In fact, let $\Omega$ and $\Omega'$ be two quantum entities associated with the Hilbert spaces $\cal H$ and ${\cal H}'$, respectively, and suppose that $\Omega$ is a subsystem of $\Omega'$, so that ${\cal H}'={\cal H} \otimes {\cal G}$ (where $\cal G$ is another Hilbert space). A (pure) state of $\Omega$ is represented by a one--dimensional projection operator $P_{\psi}$ on $\cal H$, while a (pure) state of $\Omega'$ is represented by a one--dimensional projection operator $P'_{\psi'}$ on ${\cal H}'$. Moreover, each property of the subsystem $\Omega$ is represented by a projection operator $P$ on $\cal H$ and corresponds to the property of $\Omega'$ represented by $P'=P \otimes {\mathbb {I}_{\cal G}}$ (where ${\mathbb {I}_{\cal G}}$ is the identity operator on $\cal G$). Bearing in mind our remarks at the beginning of this section and the standard results mentioned in Sec. 1, one concludes at once that the mapping $m$ in Definition 3 cannot exist in this case. Indeed, a correspondence between a pure state $p'$ of $\Omega'$ and a pure state $m(p')$ of $\Omega$ satisfying the conditions in Definition 3 can be given iff $p'$ is a product state. Thus, $\Omega$ is not a subentity of $\Omega'$ in the sense established in Definition 3, which proves our statement.

The above result shows that the standard quantum formalism cannot describe consistently physical entities made up by subentities. Yet, it seems reasonable to admit that entities of this kind exist. Hence, one meets the subentity problem: can one complete QM so that entities and subentities can be plainly described within the enlarged formalism?

According to Aerts \cite{a99a,a99b}, the subentity problem was known from the early days of QM  but it was more or less concealed by the confusion that often exists in the literature between proper and improper mixtures. Indeed, many texts and manuals on QM (see, \emph{e.g.}, \cite{j68,ctdl77}) write that, whenever a compound system is in a pure nonproduct state, the subsystems are in mixed states and not in pure states: namely, the mixed states represented by the density operators obtained by performing partial traces.  The statement that the subsystems, although they are not in a pure state, are at least in a mixed state, seems at first sight to offer a partial solution of the subentity problem in QM. Yet, this statement is incorrect, since it neglects the conceptual difference between proper and improper mixtures that we have pointed out in Sec. 1.

The solution of the subentity problem propounded by the BR approach that we want to discuss here\footnote{We note that this solution is not unique within the BR approach. An alternative solution was indeed discussed in a number of articles by Aerts and his collaborators (see, \emph{e.g.}, Ref. \cite{a00}), which is however rather conventional and does not interest us here. \label{soluzionebanale}} is rather radical and consists in interpreting improper mixtures as pure states within a more general framework, in which \emph{completed quantum entities} take the place of the quantum entities introduced above. We will not resume the BR treatment, since this would require many preliminary notions that have not been introduced in Sec. 2. Rather, we provide a new treatment that seems to us to recover the essentials of the BR proposal (the notion of completed quantum entity is however more general within the BR framework). To this end, we introduce the following definition.

\begin{definition}
Let $\Omega$ be a physical entity. We say that $\Omega$ is a \emph{completed quantum entity} iff the state property system $(\Sigma, {\cal L}, \xi, \kappa)$ describing it admits a canonical representation on a separable complex Hilbert space $\cal H$ such that:

(i) $(\Sigma, <)$ is order--isomorphic to the preordered set $({\cal W}({\cal H}), <)$ of all density operators on $\cal H$ (here $<$ denotes both the preorder on $\Sigma$ and the canonical mathematical preorder on ${\cal W}({\cal H})$ induced by inclusion of ranges);

(ii) $({\cal L}, <)$ is order--isomorphic to the complete lattice $({\cal L}({\cal H}), \subseteq)$ of all projection operators on $\cal H$;

(iii) mixed states of $\Omega$ are represented by convex sums of density operators in ${\cal W}({\cal H})$ (hence still by density operators).
\end{definition} 

By comparing Definition 4 with Theorem 1 we see that, from a standard quantum viewpoint, a completed quantum entity is characterized by a broader set of pure states with respect to a standard quantum entity, since every density operator represents a pure state.\footnote{Pure states that are represented by density operators which do not reduce to projection operators are also called \emph{density states} within the BR approach. We do not use this terminology here since we do not need to distinguish between the two kinds of pure states.} The introduction of completed quantum entities provides a plain solution of the subentity problem. Indeed, whenever a completed compound quantum entity $\Omega'$ is in the pure state $p_W$ represented by the density operator $W$, performing partial traces on $W$ produces density operators that can be interpreted as representing pure states of the component subsystems. It is then intuitively obvious, and it can be mathematically proved \cite{a99a}, that all conditions in Definition 3 are fulfilled if $\Omega$ is a component subsystem of $\Omega'$, so that $\Omega$ is a subentity of $\Omega'$ and the subentity problem disappears.

The completion of QM propounded above in order to solve the subentity problem (which can be considered a step in the search for a solution of a more general problem, the problem of generalizing QM so that also \emph{separated quantum entities} can be described by the new theory \cite{a81,a82,a02b}) puts several interesting new questions. Let us briefly sketch some of them.

(i) Let $\Omega$ be a completed quantum entity and $p_W$ a pure state of $\Omega$ represented by the density operator $W$. Then, $W$ can be expressed in many ways as a convex sum of projection operators. Yet, the coefficients that appear in the sum can never be interpreted as probabilities expressing our ignorance on the real state of $\Omega$, as it occurs when $W$ represents a proper mixture in standard QM. They can be interpreted, instead, as probabilities of actualizing particular properties, or as transition probabilities, as in the case of improper mixtures in standard QM (see Sec. 1). This enlights a shortcoming of the mathematical formalism used by the theory, namely, the Hilbert space model. Indeed, states having different physical interpretations are represented by the same mathematical objects within this formalism.

(ii) More generally, the representation of mixed states introduced in Definition 4 implies that \emph{every} mixed state is represented by a density operator on $\cal H$, which makes even deeper the problem pointed out in (i). According to Aerts, if one accepts that the set of pure states of a physical entity is represented by the set of all density operators, then this set should no longer be represented in a linear space, and this entails a breakdown of the superposition principle. Hence, the solution of the subentity problem within the BR perspective ``puts the linearity of QM at stake'' \cite{a02b}.

(iii) One can assume that the probabilistic transition from a pure to a mixed state occurring whenever a measurement is performed follows standard quantum rules (L\"uders formula), which entails that the new pure states introduced here cannot be empirically distinguished from the mixed states represented by the same density operators. This might foster the opinion that the BR solution of the subentity problem is theoretically interesting but empirically irrelevant. However, it is well known that the density operators obtained by performing partial traces in the case of a compound entity in an entangled state do not evolve unitarily. This suggests that pure states represented by density operators may evolve in a different way with respect to mixed states represented by the same density operators. Hence Aerts writes ``If we would be able to realize experimentally a nonlinear evolution of one of the subentities that has been brought into an entangled state with the other subentity as subentity of a joint entity, it would be possible to test our hypothesis and to detect experimentally whether density states are pure states or mixtures'' \cite{a00}. 

(iv) The operational justification of the standard formalism of QM summarized in Sec. 2 must be modified whenever completed quantum entities are considered. To be precise, the state property system describing an entity of this kind cannot satisfy all axioms in Definition 2, since in this case no density state could exist. Bearing in mind our comments in (ii), one is led to challenge the axiom from which linearity and superposition principle follow, namely, the covering law axiom.\footnote{It is interesting to note that if one wants a generalization of QM that can describe separated quantum entities, also the weak modularity axiom is ``put at stake'' \cite{a81,a82,a02b}.} Dealing with these topics goes, however, beyond the scopes of the present paper.     

\section{The Lecce approach}
The BR solution of the subentity problem presented in Sec. 3 is interesting but exposed to a number of objections. In particular, the introduction of states as ``modes of being'' of an entity may sound ``metaphysical'' and be irritating for many pragmatically oriented physicists. On the other hand, this notion of state is not theoretically irrelevant, since the assumption that a surjective mapping exists in Definition 3 follows from it, which could induce some scholars to reject Definition 3 and conclude that subentities in the sense specified there simply do not exist (this alternative has been seriously considered by the BR approach itself, see footnote \ref{soluzionebanale}). Therefore, as we have anticipated in Sec. 1, we want to compare the solution in Sec. 3 with another solution that can be worked out within the Lecce approach \cite{ga91}--\cite{ga05}. Hence, we summarize in this section some basic features of the Lecce approach that are needed in order to attain this goal. For the sake of brevity, we proceed by steps.

(i) A physical entity $\Omega$ is associated with a set $\Pi$ of \emph{preparing devices} and a set $\cal R$ of \emph{registering devices} which characterize the entity. A preparing device $\pi \in \Pi$ when constructed and activated, performs a \emph{preparation} of an individual sample of $\Omega$ (briefly, a \emph{physical object}). A registering device $r \in \cal R$, when constructed and activated after a preparation, performs a \emph{registration} and yields one of two possible outcomes, say \emph{yes} and \emph{no}.

(ii) Any actual setting is localized in some \emph{physical laboratory} $j$, where all preparations and all registrations can be repeated in different times. Abstractly speaking, $j$ can be considered as a \emph{space--time domain}. Let us denote the set of all laboratories by $J$.  A preparing device $\pi \in \Pi$ can be activated a great number of times in any laboratory in order to prepare \emph{ensembles} of physical objects. If every physical object in each ensemble is tested by means of a device $r \in \mathcal R$ immediately after the preparation, one obtains a frequency for the outcome yes (or no) in each ensemble. Then, we assume that preparing and registering devices are chosen in such a way that, for every pair $(\pi, r)$, these frequencies approach a limit whenever ensembles with increasing numbers of physical objects are considered, and that this limit is the same in every laboratory. 

(iii) Let $\pi_1,\pi_2 \in \Pi$ and let $\Phi_1$ and $\Phi_2$ be two ensembles of physical objects prepared by means of $\pi_1$ and $\pi_2$, respectively, in a laboratory $j$. We say that $\pi_1$ and $\pi_2$ are \emph{physically equivalent} iff every $r \in \cal R$, when applied to the physical objects in $\Phi_1$ and $\Phi_2$, yields the outcome yes with frequencies that approach the same limit (for increasing numbers of physical objects) in both ensembles and in every laboratory. This relation of physical equivalence canonically induces a partition of $\Pi$. We call \emph{states} the elements of this partition, and denote the set of all states by $\mathcal S$ in the following. We stress that this operational definition does not distinguish between pure and mixed states. 

(iv) One can consider the set of all physical objects that are prepared by means of preparing devices belonging to $\Pi$ in a laboratory $j$: this set is called the \emph{domain} $D_j$ of $j$.

(v) One can consider, in the domain $D_j$ of the laboratory $j$, the subset of all physical objects that are prepared by activating repeatedly a preparing device $\pi$: this set is the \emph{extension} $\rho_j (\pi)$ of $\pi$ in $j$. Then, the extension $\rho_j (S)$ of a state $S$ in $j$ is defined as the join of all $\rho_j (\pi)$, with $\pi \in S$.

(vi) Let $r_1,r_2 \in \cal R$ and let $\Phi_1$ and $\Phi_2$ be two ensembles of physical objects prepared by means of the same preparing device $\pi$ in a laboratory $j$. We say that $r_1$ and $r_2$ are \emph{physically equivalent} iff, when applied to the physical objects in $\Phi_1$ and $\Phi_2$, respectively, they yield the outcome yes with frequencies that approach the same limit (for increasing numbers of physical objects) in both ensembles and in every laboratory, whatever $\pi$ may be. This relation of physical equivalence canonically induces a partition of $\mathcal R$. We call \emph{effects} the elements of this partition. Within the set of all effects a subset of \emph{exact effects}, or \emph{properties}, can be selected, whose elements are classes of \emph{ideal} registering devices. We denote the set of all properties by $\mathcal E$ in the following. 

(vii) One assumes that, in the domain $D_j$ of the laboratory $j$, the subset is defined of all physical objects that would yield the outcome yes if tested by means of a given  registering device $r$ immediately after the preparation: this is the \emph{extension} $\rho_j (r)$ of $r$ in $j$. Moreover, let $r_1$ and $r_2$ be registering devices. Whenever $\rho_j (r_1)= \rho_j (r_2)$ in every laboratory $j$, $r_1$ and $r_2$ are obviously equivalent in the sense specified in (vi); yet, the converse implication does not hold \emph{a priori}. Therefore, one introduces the further assumption that \emph{$r_1$ and $r_2$ are equivalent iff, for every $j \in J$$, \rho_j (r_1)= \rho_j (r_2)$}.

(viii) By using the assumptions in (vii) one can introduce, for every $E \in \cal E$ and $j \in J$, an extension $\rho_j (E)$ of $E$ in $j$ by setting $\rho_j (E)= \rho_j (r)$, with $r \in E$. 

(ix) For every, $j \in J$, the extensions $\rho_j (S_1)$ and $\rho_j (S_2)$ of two different states $S_1$ and $S_2$ must have empty intersection, since the objects in $\rho_j (S_1)$ are prepared by devices that are  not equivalent to the devices that prepare $\rho_j (S_2)$. On the other hand, since every physical object in $D_j$ is prepared by some preparing device (see (iv)), the set of extensions of all possible states in any laboratory $j$ has to exhaust the domain $D_j$. In other words, the set $\mathcal S$ of all states induces, for every $j \in J$, a partition of $D_j$. It follows, in particular, that two states $S_1$ and $S_2$ coincide iff for every $j \in J$, $\rho_j (S_1)= \rho_j(S_2)$. Moreover, if one introduces a distinction between pure and nonpure states (see Sec. 5, (ii)) the extensions of a pure and of a nonpure state never overlap, which makes the Lecce approach essentially different from other approaches to the foundations of QM, as Ludwig's \cite{l83}.

(x) We say that a physical object in the state $S$ produces the outcome yes for every registering device in the property $E$ \emph{with certainty} iff, for every $j \in J$, $\rho_j(S) \subseteq \rho_j(E)$. Hence, one can introduce, for every $S \in \cal S$ and $E \in \cal E$, the \emph{certainly true domain} ${\mathcal E}_t(S)$ of $S$ and the \emph{certainly yes domain} ${\mathcal S}_y(E)$ of $E$, defined as follows.
\begin{eqnarray}
 {\mathcal E}_t(S)= \left\{ E \in  \mathcal E \ | \ \textrm{for every} \ j \in J, \ \rho_j(S) \subseteq \rho_j(E)  \right\}, \\
{\mathcal S}_y(E)=\left\{ S \in \mathcal S \ | \ \textrm{for every} \ j \in J, \  \rho_j(S) \subseteq \rho_j(E)  \right\}.
\end{eqnarray}
Obviously, for every $S \in \cal S$ and $E \in \cal E$, $E \in {\mathcal E}_t(S)$ iff $S \in {\mathcal S}_y(E)$. Moreover, if we put, for every $E, F \in \cal E$, $E \le F$ iff ${\mathcal S}_y(E) \subseteq {\mathcal S}_y(F)$, and for every $S, T \in {\cal S}$, $S \le T$ iff ${\mathcal E}_t(T) \subseteq {\mathcal E}_t(S)$, we get that $({\cal E}, \le)$ and $({\cal S}, \le)$ are preordered sets.

(xi) The definitions in (x) entail that an entity $\Omega$ which is characterized by the set $\cal S$, the set $\cal E$, and the mappings ${\mathcal E}_t: {\cal S} \rightarrow {\cal E}_{t}({\cal S}) \subseteq {\cal P}({\cal E})$ and ${\mathcal S}_y: {\cal E} \rightarrow {\cal S}_{y}({\cal E}) \subseteq {\cal P}({\cal S})$ (where ${\cal P}({\cal E})$ and ${\cal P}({\cal S})$ denote the power sets of $\cal E$ and $\cal S$, respectively) is a state property entity in the sense established in Sec. 2. We assume in the following that the structure $({\cal S}, {\cal E}, {\mathcal E}_t, {\mathcal S}_y)$ also is a state property system (of course, this feature of $({\cal S}, {\cal E}, {\mathcal E}_t, {\mathcal S}_y)$ can be obtained as a consequence of weaker assumptions within the Lecce approach; we do not dwell upon this subject for the sake of brevity). 

We have thus recovered within the Lecce approach a structure that is basic in the BR approach. The above framework, however, introduces (via (viii)) a typical feature of the former approach that neither occurs in the latter nor in the standard interpretation of QM, that is, \emph{objectivity of properties}. For, if an extension $\rho_j(E)$ is defined in every laboratory $j$ for every property $E$, the outcome of a registering device in $E$, when applied to a physical object $x$, does not depend on the measurement context (it is \emph{yes} iff $x \in \rho_j(E)$). In semantic terms, one can say that the truth value of a statement of the form $E(x)$ that attributes the property $E$ to the physical object $x$ is semantically defined ($E(x)$ is true iff $x \in \rho_j(E)$), independently of any theoretical or experimental procedure that may lead to know it, hence also in those physical situations in which the physical theory (QM in our case) states that it is impossible to predict it theoretically or to attain empirical knowledge of it by means of suitable measurements. In this semantic sense, which avoids any ontological commitment, properties are objective according to the Lecce approach (hence the ensuing interpretation of QM has been called \emph{Semantic Realism}, or \emph{SR}, interpretation). Of course, objectivity of properties implies that the SR interpretation clashes with the standard interpretation, which asserts instead nonobjectivity of properties on the basis of empirical (\emph{e.g.}, the double--slit experiment) or theoretical (\emph{e.g.}, the no--go theorems mentioned in footnote \ref{nogo}) arguments. Hence, the SR interpretation was worked out together with an accurate analysis of those arguments, which singled out some weaknesses in each of them \cite{ga00,gs96b,gp04}. In particular, theoretical arguments in favor of nonobjectivity turn out to be based on implicit assumptions that, when made explicit, are rather doubtful. Indeed, these assumptions subtend an epistemological perspective that assumes the validity of empirical quantum laws also in physical situations in which QM itself states that, in principle, they cannot be checked. If this perspective is criticized,\footnote{We remind that the criticism is based on a new epistemological perspective according to which the \emph{theoretical laws} of any physical theory are considered as mathematical schemes from which \emph{empirical laws} can be deduced. The latter laws are assumed to be valid in all those physical situations in which they can be experimentally checked, while no assumption of validity can be done in physical situations in which some general principle prohibits one to check them (this position is consistent, in particular, with the operational and antimetaphysical attitude of standard QM). In classical physics the new perspective does not introduce any substantial change, since there is no physical situation in which an empirical law cannot, in principle, be tested. On the contrary, if boundary, or initial, conditions are given in QM attributing noncompatible properties to the physical system (more precisely, to a sample of it), a physical situation is hypothesized that is not empirically accessible, hence no assumption of validity can be done for the empirical laws deduced from the general formalism of QM in this situation. Strangely enough, this new perspective is
sufficient to invalidate the proof the no-go theorems mentioned in footnote \ref{nogo}.} nonobjectivity of properties appears as an interpretative choice, not a logical consequence of the theory, and alternative interpretations become possible. Among these, the SR interpretation restores objectivity without requiring any change in the mathematical apparatus and in the minimal (statistical) interpretation of QM.

\section{The subentity problem within the Lecce approach}
The crucial feature of objectivity of properties within the SR interpretation (which was introduced to avoid a number of problems and paradoxes following from the standard interpretation of QM, among which, in particular, the difficulties in the quantum theory of measurement \cite{gp04}) suggests a natural solution of the subentity problem that is surprisingly similar to the solution propounded by the BR approach in a different framework. Let us discuss it proceeding again by steps.

(i) States and properties have different operational definitions (see (iii) and (vi)) and their extensions have different features (see (v), (viii) and (ix)) which implies that they must be carefully distinguished from a physical viewpoint (this distinction also occurs within the BR approach). In particular, one can never recognize the unknown state of a physical object $x$ by means of a registration procedure, just as in standard QM, where an ideal measurement \emph{puts} $x$ in a final state but provides only limited information on the initial state of $x$. On the contrary, objectivity of properties implies that it is possible to discover whether $x$ possesses or not a property $E$ by using a registering device $r \in E$ (of course, $x$ possesses $E$ iff the registration yields outcome yes), at variance with standard QM, where an ideal measurement \emph{actualizes} the measured property, that generally is neither possessed nor not possessed by $x$ before the measurement.

(ii) It follows from (i) that the probability of finding a given result when performing a measurement on a physical object $x$ can be interpreted as expressing our ignorance about the properties possessed by $x$ (in this sense one can say that it is \emph{epistemic}) within the Lecce approach, whatever the state of the physical object may be. The distinction between pure and nonpure states may still be introduced basing on the different values of the probabilities of the properties in these states, but not on different interpretations (epistemic or not) of the probabilities themselves. In particular, one can accept the standard representation of states by means of density operators, and characterize pure states as the states whose representing density operators reduce to projection operators.

(iii) We have carefully avoided to classify nonpure states as mixtures in (ii). In order to understand the reasons of this choice, let us remind that every state is operationally interpreted as an equivalence class of preparing devices in Sec. 4, (iii). If one considers a state $S$ represented by the density operator $\sum_{i} p_i |\psi_i\rangle\langle\psi_i|$, an ensemble of physical objects in the state $S$ can be realized by a \emph{mixed} preparing device, \emph{i.e.}, a device that mixes physical objects prepared by devices belonging to the states $S_1$, $S_2$, \ldots represented by the projection operators $|\psi_1\rangle\langle\psi_1|$, $|\psi_2\rangle\langle\psi_2|$, \ldots, respectively. In this case a coefficient $p_i$ cannot only be interpreted as in (ii), but also as the probability that a given physical object in the state $S$ actually is in the state $S_i$. Nevertheless, there is no evident physical reason, according to the Lecce approach, for assuming that $S$ contains only mixed preparing devices. If this assumption is avoided, referring to $S$ as a mixed state would be misleading.

(iv) It follows from (iii) that in the case of compound physical systems the density operators obtained by performing partial traces can be accepted as representing states in which also preparations occur that are not mixed in the sense specified in (iii). Whether these states must be considered ``pure'' (as in the BR approach) or ``nonpure'' (as in (iii)) is a matter of convention.

(v) One can now incorporate the BR definition of subentity within the Lecce approach by simply substituting $(\Sigma, {\cal L}, \xi, \kappa)$ and $(\Sigma', {\cal L}', \xi', \kappa')$ with  $({\cal S}, {\cal E}, {\mathcal E}_t, {\mathcal S}_y)$ and $({\cal S}', {\cal E}', {\mathcal E}'_t, {\mathcal S}'_y)$, respectively, in Definition 3. Then, let $\Omega$ and $\Omega'$ be two quantum entities associated with the Hilbert spaces $\cal H$ and ${\cal H}'$, and described by the state property systems $({\cal S}, {\cal E},  {\mathcal E}_t, {\mathcal S}_y)$ and $({\cal S}', {\cal E}', {\mathcal E}'_t, {\mathcal S}'_y)$, respectively.  Moreover, suppose that $\Omega$ is a subsystem of $\Omega'$, so that ${\cal H}'={\cal H} \otimes {\cal G}$ (where ${\cal G}$ is also a Hilbert space), and denote by $S_W$, $E_P$, $S'_{W'}$, $E'_{P'}$, the state of $\Omega$ represented by the density operator $W$ on $\cal H$, the property of $\Omega$ represented by the projection operator $P$ on $\cal H$, the state of $\Omega'$ represented by the density operator $W'$ on ${\cal H}'$ and the property of $\Omega'$ represented by the projection operator $P'$ on ${\cal H}'$, respectively. The mappings $m : S'_{W'} \in {\cal S}' \rightarrow S_W \in {\cal S}$ such that $W=Tr_{\cal G}{W'}$ (where $Tr_{\cal G}$ is the partial trace of $W'$ with respect to the subentity associated with the Hilbert space $\cal G$), and $n : E_P  \in {\cal E} \rightarrow E'_{P'} \in {\cal E}'$ such that $P'=P \otimes {\mathbb I}_{\cal G}$ (where ${\mathbb I}_{\cal G}$ is the identity in $\cal G$) can easily be proved to satisfy the conditions in Definition 3. Thus, $\Omega$ is a subentity of $\Omega'$ and the subentity problem is solved within the Lecce approach. 

The above solution can be compared with the solution proposed in Sec. 3. It is then apparent that the two solutions are compatible, apart from the convention defining the class of pure states. This result seems to us very interesting since it enhances the reliability of both solutions and may lead one to consider more carefully the proposal of broadening standard QM that comes out from the BR and the Lecce approach.

It remains to stress that the probabilistic definition of states in Sec. 4, (iii) (which is usual in the literature on the foundations of QM and not specific of the Lecce approach \cite{bc81}) groups together, in the case of nonpure states, mixed with nonmixed preparing devices, that therefore cannot be distinguished by means of measurements. This explains the deep roots of the indistinguishability problem mentioned in Sec. 3, (iii). Moreover, it opens the way to a possible solution of the problem of explaining how both unitary and nonunitary evolutions may occur for the same density operator (see again, Sec. 3, (iii)). Indeed, it suggests distinguishing mixed from nonmixed preparing devices by introducing a new equivalence relation on $\Pi$, strictly contained in the physical equivalence relation defined in Sec. 4, (iii). Thus every state $S$ would be associated with a family of \emph{hidden states} (which seemingly introduces a kind of hidden variables theory; there are however some important peculiarities that we cannot discuss here, see, \emph{e.g.}, \cite{ga05}). These would be equivalent with respect to measurements but could have different behaviours with respect to time evolution.

 The above suggestion seems especially suitable for concluding this paper. Indeed, it provides a natural support to Aerts' proposals in Sec. 3, (iii), which again shows similarities between the BR and the Lecce approaches, notwithstanding their remarkable differences.

\end{document}